\documentclass[twocolumn,showpacs,aps,prl,superscriptaddress,floatfix]{revtex4}
\usepackage{graphicx}
\usepackage{verbatim}
\usepackage{dcolumn}
\usepackage{amsmath}
\usepackage{epsfig}
\usepackage{subfigure}

\newcommand{\BaBarYear}      {16}
\newcommand{\BaBarNumber}    {003}
\newcommand{\BaBarType}      {PUB}  
\newcommand{\SLACPubNumber}  {16549}

\input babarsym.tex

\def\epem{e^+e^-}
\def\mpmm{\mu^+\mu^-}

\def\pipm{\pi^+\pi^-}

\def\figurebox#1#2#3{%
    \def\arg{#3}%
    \ifx\arg\empty
    {\hfill\vbox{\hsize#2\hrule\hbox to #2{\vrule\hfill\vbox to #1{\hsize#2\vfill}\vrule}\hrule}\hfill}%
    \else
    {\hfill\epsfbox{#3}\hfill}%
    \fi}

\begin{document}

\pagestyle{plain}

\begin{flushleft}
\babar-\BaBarType-\BaBarYear/\BaBarNumber \\
SLAC-PUB-\SLACPubNumber\\
\end{flushleft}

\title{{\large \bf Search for a muonic dark force at \babar}}

%
%
\author{J.~P.~Lees}
\author{V.~Poireau}
\author{V.~Tisserand}
\affiliation{Laboratoire d'Annecy-le-Vieux de Physique des Particules (LAPP), Universit\'e de Savoie, CNRS/IN2P3,  F-74941 Annecy-Le-Vieux, France}
\author{E.~Grauges}
\affiliation{Universitat de Barcelona, Facultat de Fisica, Departament ECM, E-08028 Barcelona, Spain }
\author{A.~Palano}
\affiliation{INFN Sezione di Bari and Dipartimento di Fisica, Universit\`a di Bari, I-70126 Bari, Italy }
\author{G.~Eigen}
\affiliation{University of Bergen, Institute of Physics, N-5007 Bergen, Norway }
\author{D.~N.~Brown}
\author{Yu.~G.~Kolomensky}
\affiliation{Lawrence Berkeley National Laboratory and University of California, Berkeley, California 94720, USA }
\author{H.~Koch}
\author{T.~Schroeder}
\affiliation{Ruhr Universit\"at Bochum, Institut f\"ur Experimentalphysik 1, D-44780 Bochum, Germany }
% ...
\collaboration{The \babar\ Collaboration}
\noaffiliation

\begin{abstract}
Many models of physics beyond the Standard Model predict the existence of new Abelian forces with new 
gauge bosons mediating interactions between ``dark sectors'' and the Standard Model. We report a search 
for a dark boson $Z'$ coupling only to the second and third generations of leptons in the reaction $\epem \rightarrow \mpmm Z', 
Z' \rightarrow \mpmm$ using $514 \fb^{-1}$ of data collected by the \babar\ experiment. No significant signal is 
observed for $Z'$ masses in the range $0.212-10\gev$. Limits on the coupling parameter $g'$ as low as 
$7\times 10^{-4}$ are derived, leading to improvements in the bounds compared to those previously 
derived from neutrino experiments. 
\end{abstract}


\maketitle

\setcounter{footnote}{0}
In spite of the many successes of the Standard Model (SM), the known particles and interactions are 
insufficient to explain cosmological and astrophysical observations of dark matter. This motivates the possibility of new 
hidden sectors that are only feebly coupled to the SM; by analogy with the SM, such sectors may contain their own 
interactions with new gauge bosons ($Z'$). In the simplest case of a hidden $\mathrm{U}(1)$ interaction, SM fields may 
directly couple to the $Z'$, or alternatively the $Z'$ boson may mix with the SM hypercharge boson, which typically 
results from an off-diagonal kinetic term~\cite{Holdom:1985ag}. In the latter case, the $Z'$ inherits couplings 
proportional to the SM gauge couplings; due to large couplings to electrons and light-flavor quarks, such scenarios 
are strongly constrained by existing searches~\cite{Essig:2013lka,Babusci:2014sta,Merkel:2014avp,Lees:2014xha,Batley:2015lha,
Anastasi:2015qla, Anastasi:2016lwm}. 

When SM fields are directly charged under the dark force, however, the $Z'$ may interact preferentially with heavy-flavor 
leptons, greatly reducing the sensitivity of current searches. Such interactions could account for the experimentally 
measured value of the muon anomalous magnetic dipole moment~\cite{Pospelov:2008zw}, as well as the discrepancy in 
the proton radius extracted from measurements of the Lamb shift in muonic hydrogen compared to observations in non-muonic 
atoms~\cite{Barger:2010aj,TuckerSmith:2010ra}. Direct $Z'$ couplings to left-handed leptons also lead to new interactions 
involving SM neutrinos that increase the cosmological abundance of sterile neutrinos mixing with SM neutrinos, consistent 
with the observed dark matter abundance~\cite{Shuve:2014doa}.

We report herein a search for dark bosons $Z'$ with vector couplings only to the second and third generations of leptons~\cite{He:1990pn,He:1991qd} 
in the reaction $\epem \rightarrow \mpmm Z', Z' \rightarrow \mpmm$. While such a scenario can be additionally constrained by neutrino-nucleus 
scattering at neutrino beam experiments, the measurement presented here is also sensitive to models where couplings to neutrinos are absent, 
such as a gauge boson coupled exclusively to right-handed muons~\cite{Batell:2011qq}. This search is based on 514 fb$^{-1}$ of data collected 
by the \babar\ detector at the PEP-II $\epem$ storage ring, mostly taken at the $\Y4S$ resonance, but also at the $\Y3S$ and $\Y2S$ 
peaks, as well as in the vicinity of these resonances~\cite{Lees:2013rw}. The \babar\ detector is described in detail 
elsewhere~\cite{Bib:Babar,TheBABAR:2013jta}. Dark boson masses between the dimuon threshold and $10 \gev$ are 
probed~\cite{units}. To avoid experimental bias, the data are only examined after finalizing the analysis strategy. A sample 
of about 5\% of the dataset is used to optimize and validate the analysis strategy, and is then discarded.

Signal events are simulated by MadGraph 5~\cite{Alwall:2014hca} and hadronized in Pythia 6~\cite{Sjostrand:2006za} for $Z'$ mass 
hypotheses ranging from the dimuon mass threshold to $10.3 \gev$. The background arises mainly from QED processes. The 
$\epem \rightarrow \mpmm \mpmm$ reaction is generated with Diag36~\cite{Berends:1984gf}, which includes the full set of lowest 
order diagrams, while the $\epem \rightarrow \mpmm (\gamma)$ and $\epem \rightarrow \tau^+ \tau^- (\gamma)$ processes are 
simulated with KK~\cite{Jadach:2000ir}. Other sources of background include $\epem \rightarrow q\overline{q}$ ($q=u,d,s,c$) 
continuum production, simulated with JETSET~\cite{Sjostrand:1993yb}, and $\epem \rightarrow \pipm \jpsi$ events, generated 
using EvtGen~\cite{Lange:2001uf} with a phase-space model. The detector acceptance and reconstruction efficiencies are determined 
using a Monte Carlo (MC) simulation based on GEANT4~\cite{Bib::Geant}. 

We select events containing two pairs of oppositely-charged tracks, where both positively-charged or both negatively-charged tracks
are identified as muons by particle identification algorithms (PID). Identifying only two muons maintains high signal efficiency while 
rejecting almost all background sources but $\epem \rightarrow \mpmm\mpmm$ events. In addition, the sum of energies of 
electromagnetic clusters above $30 \mev$ not associated to any track must be less than 200 MeV to remove background containing neutral 
particles. To suppress background from the decay chain $\Upsilon(3S,2S) \rightarrow \pi^+ \pi^- \Y1S, \Y1S \rightarrow \mpmm$, we 
reject events taken on the $\Y2S$ or $\Y3S$ peaks containing any pair of oppositely charged tracks with any dimuon invariant mass 
within $100 \mev$ of the nominal $\Y1S$ mass.

\begin{figure}[htb]
\begin{center}
  \includegraphics[width=0.48\textwidth]{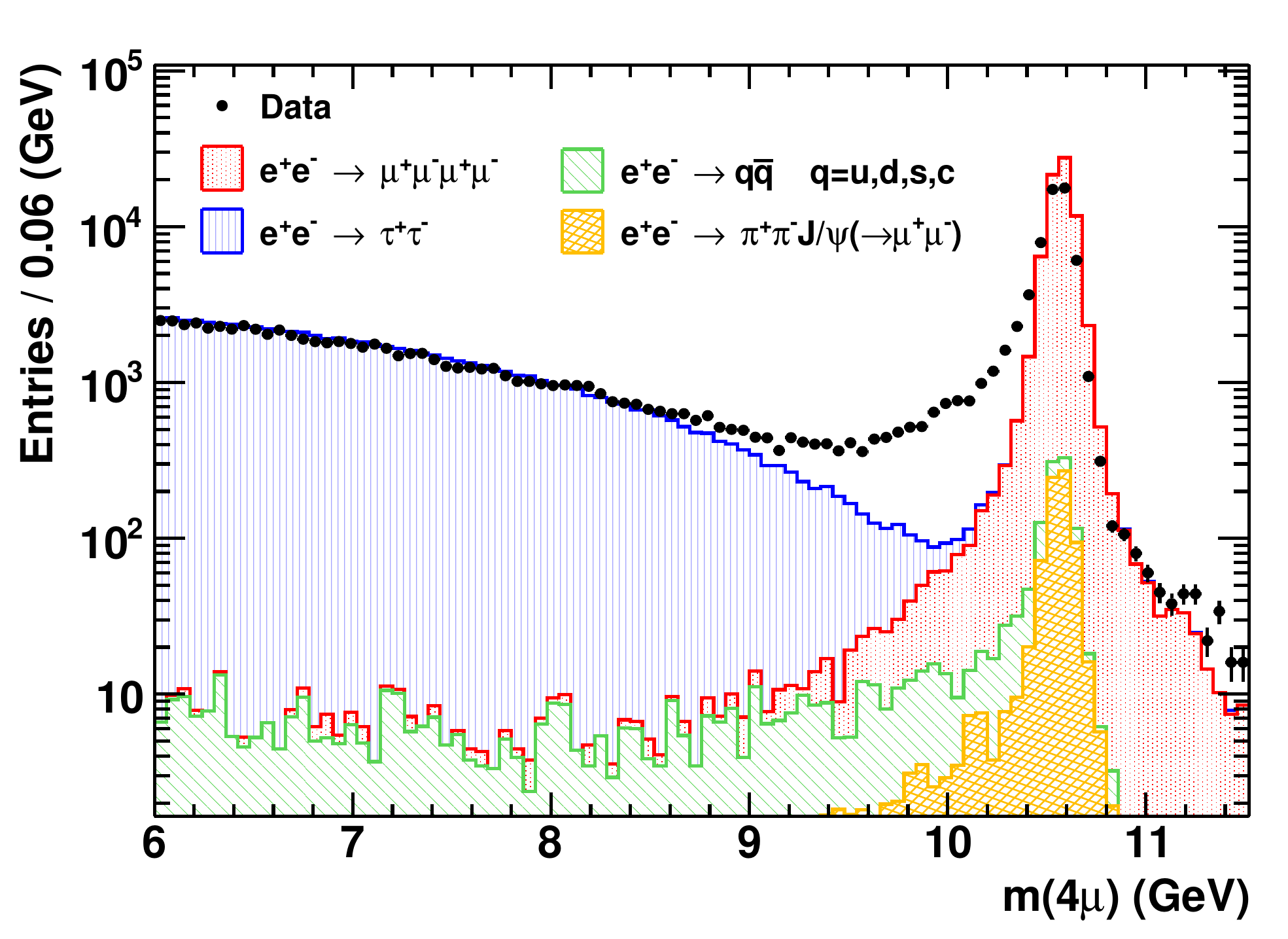}
\end{center}
\caption
{The distribution of the four-muon invariant mass, $m(4\mu)$, for data taken at the $\Y4S$ peak together with Monte Carlo 
predictions of various processes normalized to data luminosity. The $\epem \rightarrow \mpmm \mpmm$ Monte Carlo does not 
include ISR corrections.}
\label{Fig1}
\end{figure}

The distribution of the four-muon invariant mass after these cuts is shown in Fig.~\ref{Fig1} for the data taken at the 
$\Y4S$ center-of-mass (CM) energy. The background at low masses is fairly well reproduced by the simulation, while the 
$\epem \rightarrow \mpmm \mpmm$ Monte Carlo overestimates the full-energy peak by $\sim 30\%$ and fails to reproduce the 
radiative tail. This is expected, since Diag36 does not simulate initial state radiation (ISR). We further select 
$\epem \rightarrow \mpmm \mpmm$ events by requiring a four-muon invariant mass within $500 \mev$ of the nominal CM energy, 
allowing for the possibility of ISR emission. The four-muon system is finally fitted, constraining its CM energy to be within 
the beam energy spread and the tracks to originate from the interaction point to within its uncertainty. This kinematic fit is 
solely used to improve the $Z'$ mass resolution of the bulk of events near the full-energy peak; no further requirement is imposed 
on the fit quality. We do not attempt to select a single $Z' \rightarrow \mpmm$ candidate per event, but simply consider all 
combinations. 

\begin{figure}[htb]
\begin{center}
  \includegraphics[width=0.48\textwidth]{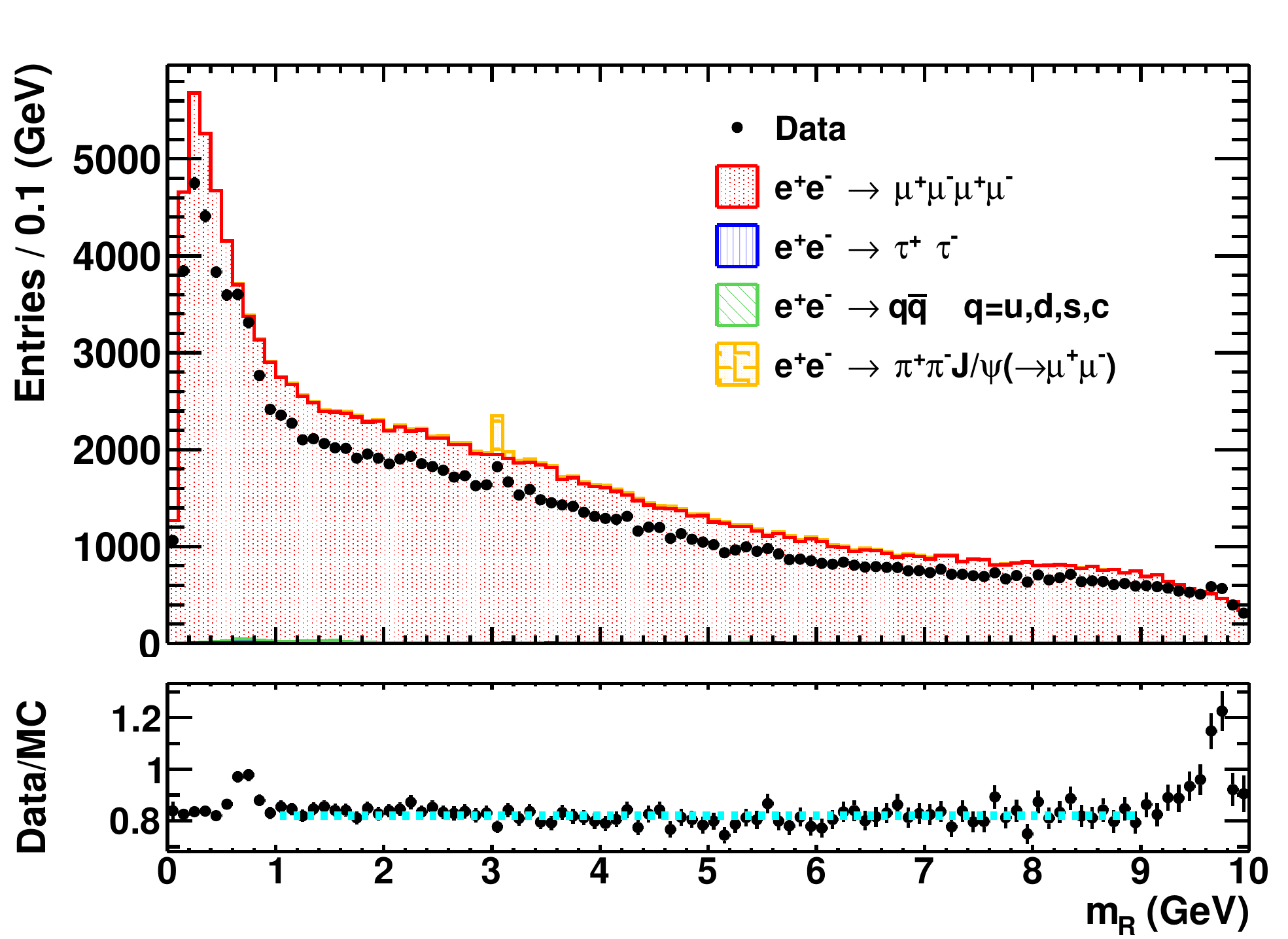}
\end{center}
\caption
{The distribution of the reduced dimuon mass, $m_R$, together with Monte Carlo predictions of various processes 
normalized to data luminosity. Four combinations per event are included. The fit of the ratio between reconstructed 
and simulated events is shown as a light blue dashed line. The $\epem \rightarrow \mpmm \mpmm$ Monte Carlo does not 
include ISR or other efficiency corrections (see text).}
\label{Fig2}
\end{figure}

The distribution of the reduced dimuon mass, $m_R = \sqrt{m_{\mpmm}^2 -4m_\mu^2}$, is shown in Fig.~\ref{Fig2}, 
together with the predictions of various Monte Carlo simulations. The reduced  mass has a smoother behavior near threshold 
and is easier to model than the dimuon mass. The spectrum is dominated by $\epem \rightarrow \mpmm \mpmm$ production, 
with additional contributions from $\epem \rightarrow \pipm \rho, \rho \rightarrow \pipm$, $\epem \rightarrow \mu^+ \mu^- 
\rho, \rho \rightarrow \pipm$, and $\epem \rightarrow  \pipm J/\psi, J/\psi \rightarrow \mu^+ \mu^-$ events, where one or 
several pions are misidentified as muons. A peak corresponding to the $\rho$ meson is visible at low mass; the second 
$Z'$ candidate reconstructed in these events generates the enhancement near $9.5 \gev$. Other than the $\jpsi$, no significant 
signal of other narrow resonances is observed.

The signal efficiency rises  from $\sim 35\%$ at low masses to $\sim 50\%$ around $m_R = 6-7 \gev$, before 
dropping again at higher masses. The signal efficiencies include a correction factor of 0.82, which primarily accounts 
for the impact of ISR not included in the simulation, as well as differences between data and simulation in 
trigger efficiency, charged particle identification, and track and photon reconstruction efficiencies. This correction 
factor is derived from the ratio of the $m_R$ distribution in simulated $\epem \rightarrow \mpmm \mpmm$ events to 
the observed distribution in the mass region 1--9 $\gev$, excluding the $\jpsi$ region (light blue line in Fig. 2). 
An uncertainty of 5\% is propagated as a systematic uncertainty, covering the small variations between data-taking 
periods and the uncertainties on the $\epem \rightarrow \mpmm \mpmm$ cross-section. 

We extract the signal yield as a function of $m_{Z'}$ by performing a series of unbinned maximum likelihood fits to the reduced 
dimuon mass spectrum, covering the mass range $m_{R} < 10 \gev$ for the data taken near the $\Y4S$ resonance, and up to $9 \gev$ 
for the datasets collected near the $\Y2S$ and $\Y3S$ resonances. The search is conducted in varying mass steps that correspond 
to the dark boson mass resolution. Each fit is performed over an interval 50 times broader than the signal resolution at that 
mass for $m_R > 0.2 \gev$, or over a fixed interval $0 - 0.3 \gev$ for $m_R < 0.2 \gev$. We estimate the signal resolution by 
Gaussian fits to several simulated $Z'$ samples for the purpose of determining the scan steps, and interpolate the results to 
all other masses. The resolution varies between $1-9\mev$, dominated by experimental effects. We probe a total of 2219 mass 
hypotheses. The bias in the fitted values, estimated from a large ensemble of pseudo-experiments, is negligible.

The likelihood function, described below, contains components from signal, continuum background, and peaking 
background where appropriate. The signal probability density function (pdf) is modeled directly from the 
signal Monte Carlo mass distribution using a non-parametric kernel density function. The pdf is interpolated 
between the known simulated masses using an algorithm based on the cumulative density function~\cite{Read:1999kh}. 
An uncertainty of $0.1-3.2$ events associated to this procedure is estimated by taking the next-to-closest mass point
in place of the closest simulated mass point to interpolate the signal shape. The agreement between the simulated 
signal resolution and the data is assessed by fitting the full-energy peak of the four-muon invariant mass 
spectrum in the range $10.3 - 10.7 \gev$ with a Crystal Ball function~\cite{CrystalBall}. The ratio of simulated and reconstructed peak 
widths is $1.01 \pm 0.04$, consistent with unity. The impact of ISR emission on the peak widths are expected to be 
small in that mass range. Similarly, the decay width of the $J/\psi$ resonance is well reproduced by the simulation 
within its uncertainty.

The background is described by a function of the form $\arctan(ax+bx^2+cx^3)$ for fits in the low mass region, 
where $a,b,c$ are free parameters, and by a second order polynomial above $m_R = 0.2 \gev$. The two methods give similar 
signal yields at the transition point. Peaking contributions from the $\jpsi$ 
resonance are modeled from the mass distribution extracted from the corresponding Monte Carlo, leaving the yield as a
free parameter. We exclude the resonant region from the search, vetoing a range of $\pm 30 \mev$ around the nominal $\jpsi$ 
mass. The contribution from $\rho$-meson decay is very wide and easily absorbed by the background fit in each narrow 
window. We estimate the uncertainty associated with the background model by repeating the fit using a third order polynomial 
in the high-mass region or a fourth-order polynomial constrained to pass through the origin in the low mass range. 
This uncertainty is as large as 35\% of the statistical uncertainty in the vicinity of the dimuon threshold 
and high-mass boundary, but remains at a level of a few percent outside these regions. 

\begin{figure}
\begin{center}
  \includegraphics[width=0.48\textwidth]{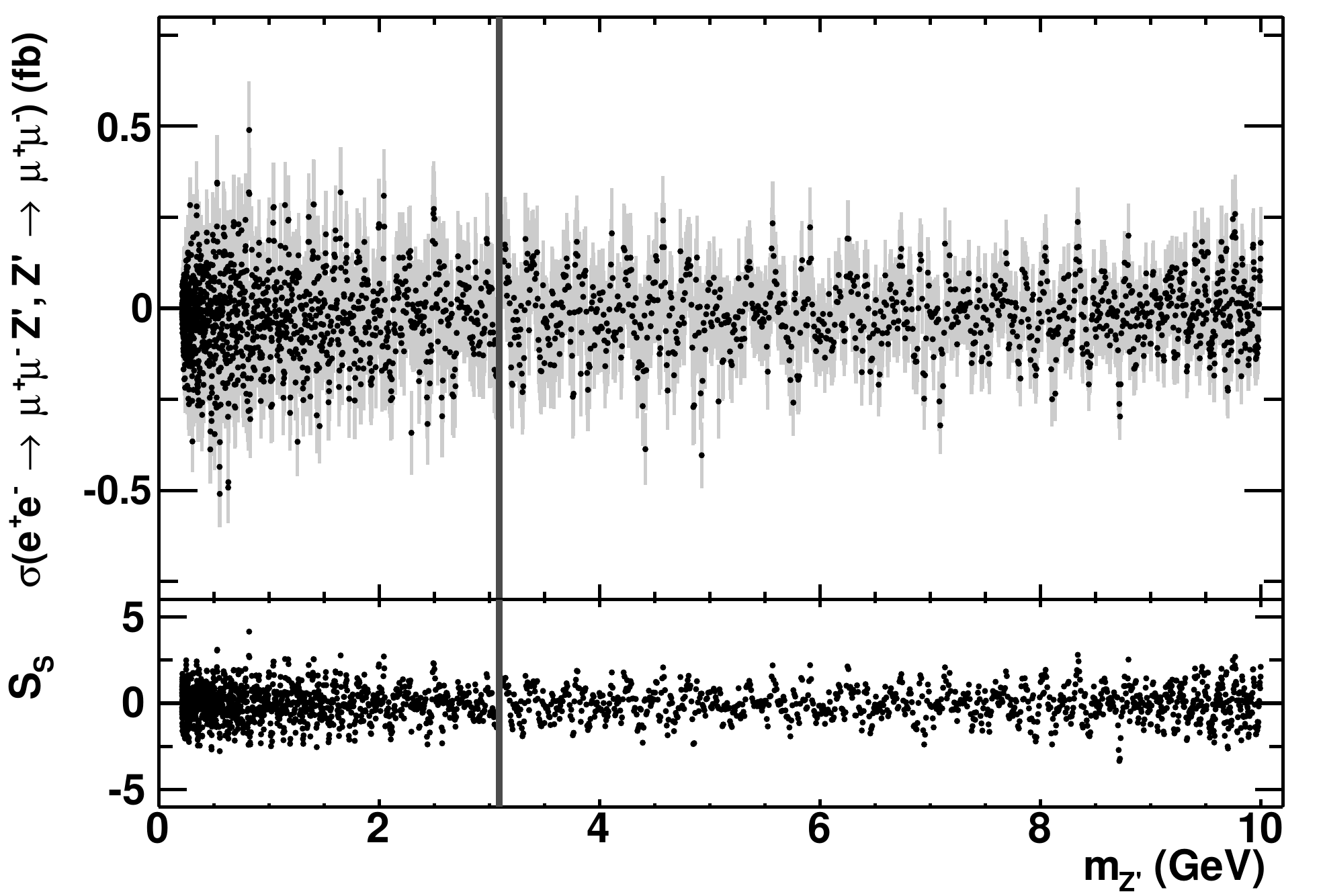}
\caption{The measured $\epem \rightarrow \mpmm Z', Z' \rightarrow \mpmm$ cross-section together with its statistical 
significance, $S_S$ (see text for definition), as a function of the $Z'$ mass. The uncertainty on each point is 
shown as light gray error bars. The dark gray band indicates the region excluded from the analysis.}
\label{fig3}
\end{center}
\end{figure}

The $\epem \rightarrow \mpmm Z', Z' \rightarrow \mpmm$ cross-section is extracted for each dataset as a function 
of the $Z'$ mass by dividing the signal yield by the efficiency and luminosity. The uncertainties on the luminosity 
(0.6\%)~\cite{Lees:2013rw} and the limited Monte Carlo statistics (1--3\%) are propagated as systematic 
uncertainties. The cross-sections are finally combined and displayed in Fig.~\ref{fig3}. We consider all but the 
uncertainties on the luminosity and the efficiency corrections to be uncorrelated. 
The statistical significance of each fit is taken as ${\cal S_S} = {\rm sign}(N_{\rm sig})\sqrt{2\log{({\cal L / L}_0})}$, 
where $N_{\rm sig}$ is the fitted signal yield, and $\cal L$ (${\cal L}_0$) is the maximum likelihood values for a fit 
including (excluding) a signal. These significances are almost Gaussian, and the combined significance is derived under 
this assumption. A large sample of Monte Carlo experiments is generated to estimate trial factors. The largest local 
significance is $4.3\sigma$, observed near $m_{Z'} = 0.82\gev$, corresponding to a global significance of $1.6\sigma$, 
consistent with the null hypothesis. 

\begin{figure}
\begin{center}
 \includegraphics[width=0.48\textwidth]{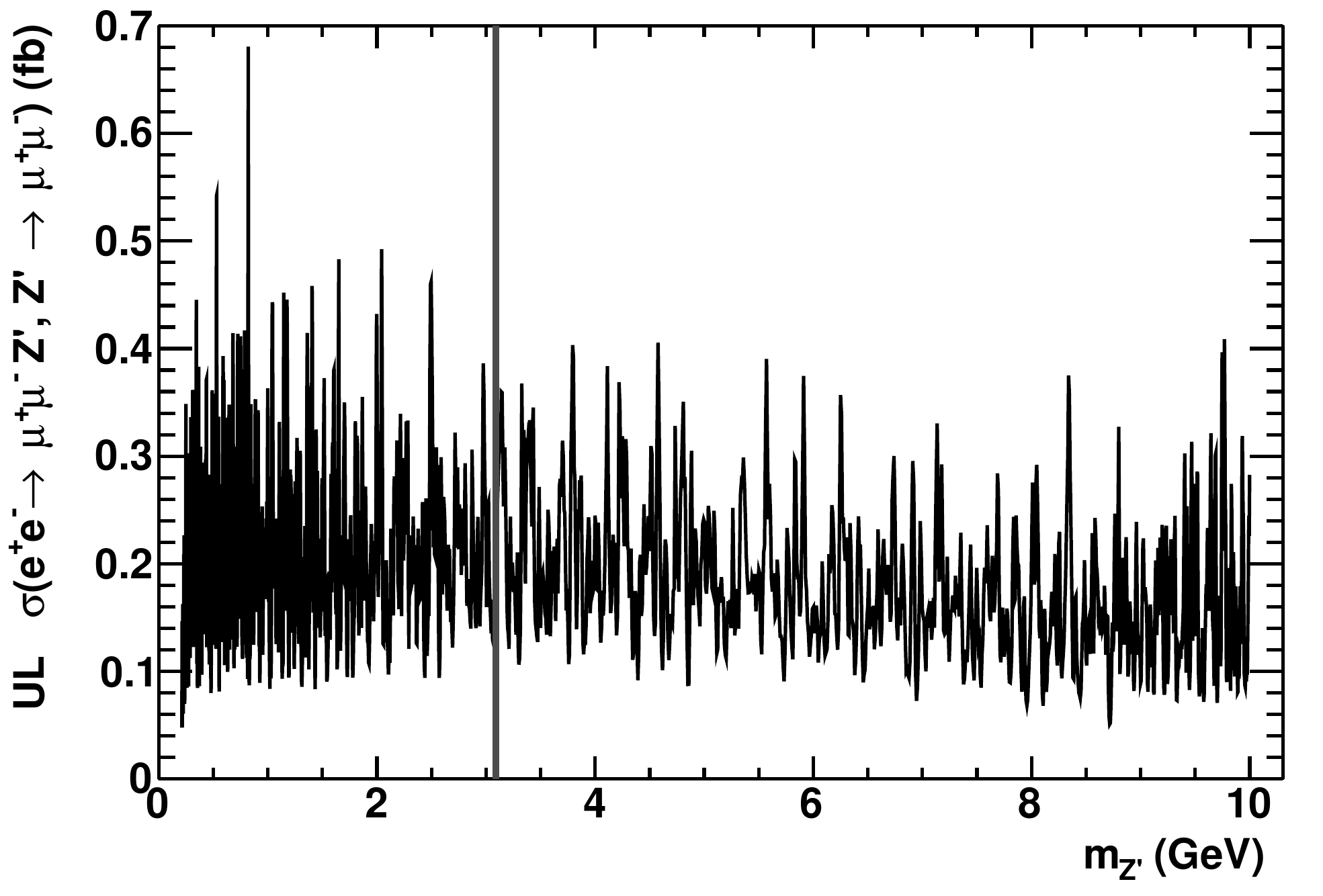}
\caption{The 90\% CL upper limits on the cross-section $\sigma(\epem \rightarrow \mpmm Z', Z' \rightarrow \mpmm)$ as a function 
of the $Z'$ mass. The dark gray band indicates the region excluded from the analysis.}
\label{fig4}
\end{center}
\end{figure}

\begin{figure}
\begin{center}
 \includegraphics[width=0.48\textwidth]{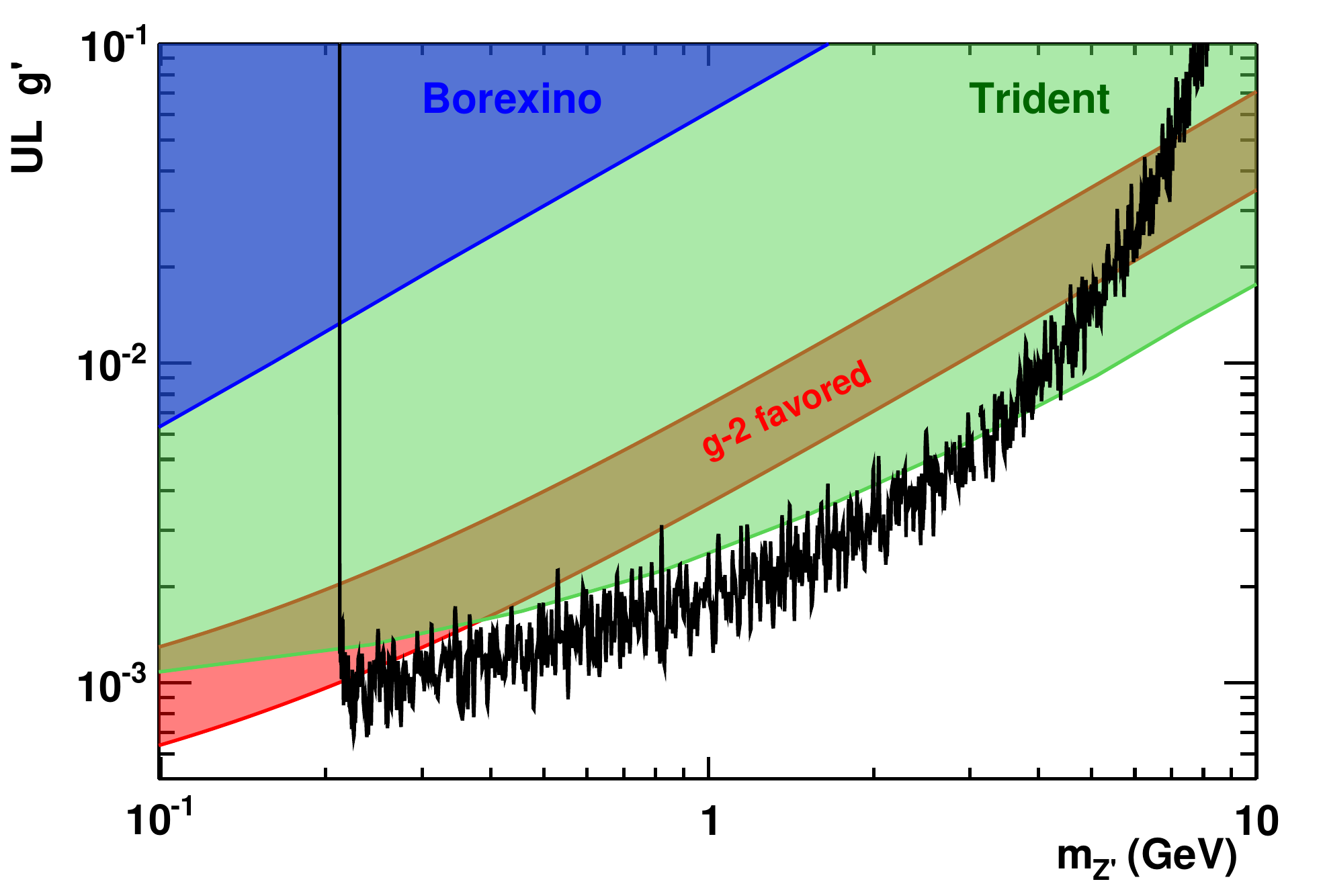}
\caption{The 90\% CL upper limits on the new gauge coupling $g'$ as a function of the $Z'$ mass, together with the 
constraints derived from the production of a $\mpmm$pair in $\nu_\mu$ scattering (``Trident" 
production) ~\protect\cite{Altmannshofer:2014pba,Kamada:2015era}. The region consistent with the discrepancy between 
the calculated and measured anomalous magnetic moment of the muon within $2\sigma$ is shaded in red.}
\label{fig5}
\end{center}
\end{figure}

We derive 90\% confidence level (CL) Bayesian upper limits (UL) on the cross-section $\sigma(\epem \rightarrow \mpmm Z', Z' \rightarrow \mpmm)$, 
assuming a uniform prior in the cross-section by integrating the likelihood from zero up to 90\% of its area. Correlated (uncorrelated) 
systematic uncertainties are included by convolving the combined (individual) likelihood with Gaussian distributions 
having variances equal to the corresponding uncertainties. The results are displayed in Fig.~\ref{fig4} as a function 
of the $Z'$ mass. The corresponding 90\% CL upper limits on the coupling parameter $g'$ in the scenario with equal magnitude 
vector couplings to muons, taus, and the corresponding neutrinos are shown in Fig.~\ref{fig5}, together 
with constraints derived from neutrino experiments~\cite{Altmannshofer:2014pba}. Upper limits down to $7\times10^{-4}$ near 
the dimuon threshold are set. 

In summary, we report the first search for the direct production of a new muonic dark force boson, providing a model-independent test 
of theories with new light particles coupled to muons. For identical coupling strength to muons, taus, and the corresponding neutrinos, 
we exclude all but a sliver of the remaining parameter space preferred by the discrepancy between the calculated and measured anomalous 
magnetic moment of the muon above the dimuon threshold~\cite{Altmannshofer:2014pba}, and we set the strongest bounds for nearly all of the 
parameter space below $\sim 3 \gev$. Because this search relies only on the $Z'$ coupling to muons, the result can also be 
interpreted giving powerful constraints on other new vectors and scalars that interact exclusively with muons.

\section{Acknowledgments}
We thank Maxim Pospelov for helpful conversations.
We are grateful for the 
extraordinary contributions of our \pep2\ colleagues in
achieving the excellent luminosity and machine conditions
that have made this work possible.
The success of this project also relies critically on the 
expertise and dedication of the computing organizations that 
support \babar.
The collaborating institutions wish to thank 
SLAC for its support and the kind hospitality extended to them. 
This work is supported by the
US Department of Energy
and National Science Foundation, the
Natural Sciences and Engineering Research Council (Canada),
the Commissariat \`a l'Energie Atomique and
Institut National de Physique Nucl\'eaire et de Physique des Particules
(France), the
Bundesministerium f\"ur Bildung und Forschung and
Deutsche Forschungsgemeinschaft
(Germany), the
Istituto Nazionale di Fisica Nucleare (Italy),
the Foundation for Fundamental Research on Matter (The Netherlands),
the Research Council of Norway, the
Ministry of Education and Science of the Russian Federation, 
Ministerio de Econom\'{\i}a y Competitividad (Spain), the
Science and Technology Facilities Council (United Kingdom),
and the Binational Science Foundation (U.S.-Israel).
Individuals have received support from 
the Marie-Curie IEF program (European Union) and the A. P. Sloan Foundation (USA). 



\begin{thebibliography}{9}
\bibitem{units} Natural units ($\hbar=c=1$) are used throughout this paper.
\end{thebibliography}
\end{document}